\documentclass[preprintnumbers,amsmath,11pt,amssymb,floatfix,superscriptaddress,nofootinbib]{article}

\def\mytitle#1{\setcounter{equation}{0}
\setcounter{footnote}{0}
\begin{flushleft}\Large\textbf{#1}\end{flushleft}
\vspace{0.25cm}}
\def\myname#1{\leftline{{\large #1}}\vspace{-0.13cm}}
\def\myplace#1#2{\small\begin{flushleft}\textit{#1}\\
\texttt{#2}\end{flushleft}}

\def\myclassification#1{\small\noindent
Keywords :
       #1\vspace{0.5cm}}
\usepackage{graphicx}
\begin{document}
\mytitle{Thermodynamic Studies of Different Black Holes with Modifications of Entropy}

\myname{$Amritendu~ Haldar^{*}$\footnote{amritendu.h@gmail.com} and $Ritabrata~
Biswas^{\dag}$\footnote{biswas.ritabrata@gmail.com}}
\myplace{*Department of Physics, Sripat Singh College, Jiaganj, Murshidabad $-$ 742123, India.\\$\dag$ Department of Mathematics, The University of Burdwan, Golapbag Academic Complex, City : Burdwan  $-$ 713104, Dist. : Purba Barddhaman, State : West Bengal, India.} {}
 
\begin{abstract}
In recent years, the thermodynamic properties of black holes are topics of interests. We investigate the thermodynamic properties like surface gravity and Hawking temperature on event horizon of regular black holes viz. {\it Hayward Class} and {\it asymptotically AdS (Anti-de Sitter)} black holes. We also analyze the {\it thermodynamic volume} and {\it naive geometric volume} of  asymptotically AdS black holes and show that the entropy of these black holes is simply the ratio of the naive geometric volume to thermodynamic volume. We plot the different graphs and interpret them physically. We derive the {\it `cosmic-Censorship-Inequality'} for both type of black holes. Moreover, we calculate {\it the thermal heat capacity} of aforesaid black holes and study their stabilities in different regimes. Finally, we compute the logarithmic correction to the entropy for both the black holes considering the quantum fluctuations around the thermal equilibrium and study the corresponding thermodynamics.         
   
\end{abstract}

\myclassification{Black Hole Temperature; Statistical Quantum Fluctuations; Black Hole Entropy; Thermal Stability and Gravitational Mass.}

\section{Introduction}
The most interesting study regarding the compact objects like Black Holes (BHs) is that they can be treated as a thermodynamic system. This idea came forward with the proposition of Hawking, S. W. \cite{Bardeen 1973, Hawkings 1974} who has proposed the model that BHs can radiate via quantum tunnelling. This phenomena is famous as Hawking radiation. The best thing regarding the BHs is the absence of any hard surface and they have an exact geometrical boundary named as event horizon. The thermodynamics of BHs mainly comprises of the studies of the analogous thermodynamic parameters which are defined on the event horizon. The entropy of BHs is represented by surface area at event horizon (as both entropy and surface area are ever increasing parameters). On the other hand, the surface gravity which is non negative ever is related to the Hawking temperature. To make the scenario of the analogy more clear, consider a quasi-static process during which a stationary BH of mass $M$, angular momentum $J$ and surface area $A$ is taken to a new stationary BH with parameters $M + \delta M$, $J+ \delta J$ and $A + \delta A $. This perturbation is done by the addition of a small quantity of matter described by the stress-energy tensor $T_{\alpha \beta}$. Mass and angular momentum of the BH increase by amounts:
	$\delta M = - \int_H T^\alpha_\beta \xi^\beta_{(t)} dH_\alpha $ and 
	$\delta J = - \int_H T^\alpha_\beta \xi^\beta_{(\phi)} dH_\alpha ,$ where
	$\xi^{\beta}_{(t)}$ is the Killing vector for the translation and 
	 $\xi^\beta_{(\phi)}$ is the rotational Killing vector.
	 The integration is over the entire event horizon $H$   and $dH_\alpha = - \xi_\alpha dA dv$, is the surface element on ${H}$,   with co-ordinates $y = \left(v, \theta^{A}\right)$. $\xi^\alpha = \xi^{\alpha}_{(t)} + \Omega_H \xi^\alpha_{(\phi)}$,   where $\Omega_H$ is the angular velocity of the BH. All these definitions bring us to,
	$\delta M - \Omega_H \delta J = \int_H T_{\alpha \beta} \left(\xi^\beta_{(t)} + \Omega_H \xi^\beta_{(\phi)}\right) \xi^\alpha dA ~dv = \int dv \oint_H T_{\alpha\beta} \xi^{\alpha} \xi^{\beta} dA$. Now a simplified Raychoudhuri's equation \cite{Poisson 2004}  for terms like expansion $\theta$ and shear $\sigma_{\alpha \beta}$ are of first order in $T_{\alpha \beta}$. Neglecting quadratic terms, we have $\frac{d\theta}{dv} = \kappa \theta - 8 \pi T_{\alpha \beta} \xi^\alpha \xi^\beta$. Both before and after the perturbation, the BH is stationary, i.e., $\theta(v = \pm \infty) = 0$,   we obtain, $\delta M - \Omega_H \delta J$ $ = - \frac{1}{8 \pi} \int dv \oint_H \left(\frac{d\theta}{d v} - \kappa \theta\right) dA$ $ = \frac{\kappa}{8 \pi} \int dv \oint_H \theta dA$, where $\theta$ is the fractional rate of change of the congruence's cross-sectional area $\Rightarrow ~~ $ $\delta M - \Omega_H \delta J $ $=\frac{\kappa}{8 \pi} \int dv \oint_H \left(\frac{1}{dA} \frac{d}{dv} dA\right) dA =\frac{\kappa}{8 \pi} \oint_H dA $ $=\frac{\kappa}{8 \pi} \delta A,$ which is the statement of first law of BH thermodynamics\cite{Bardeen 1973, Bekenstein 1973}. Hawking and Page proposed in their seminal paper \cite{Page 1983} that there exist a phase transition in the phase spaces between Schwarzschild-AdS BH and thermal radiation. This phase transition is called Hawking-Page phase transition which can be explained as the confinement/deconfinement of phase transition of gauge field in the AdS/CFT corresponds \cite{Maldacena 1998, biswas1, biswas2, biswas3, biswas4, biswas5}.
	
Area theorem gives rise to the second law of BH mechanics : If the null energy condition is satisfied, the surface area of the BH can never decrease \cite{Hawkings 1971}. A physical realization  of the entropy may be done by considering a quantity of thermal radiation and following what entropy it would have when it forms a BH. In Planck units, i.e.,  $G =c = \hbar = \kappa = 1$,  a ball of radiation at temperature $T$ and radius $R$ has mass $M \sim T^4 R^3$, and entropy $S \sim T^3R^3$.
The radiation will form a BH when $R \sim M$ which implies $T \sim M^{-\frac{1}{2}}$ and hence $S \sim M^\frac{3}{2}$.
This signifies the entropy, $S_{BH} \sim M^2$. So any BH with $M>>1$ (mass $\gg$ Planck mass), has an entropy much larger than the entropy of the thermal radiation that formed it. On the basis of classical general relativity and considering simple dimensional analysis, Bardeen, Carter and Hawking (BCH) showed that BH in equilibrium obeys two basic laws of thermodynamics, viz. zeroth law and first law of thermodynamics and BH temperature $ T $, surface gravity $ \kappa $ and area $ A $ of BH are related to each other \cite{Pradhan 2016a, Fan 2016}. The calculation of specific heat of BH is very important to study the stability criteria of BHs. The BHs having positive specific heat are thermally stable where as the BHs having negative specific heat are thermally unstable.

While deriving a BH solution from Einstein's equation if we consider the Einstein tensor to depend on cosmological constant, the ultimate solution inherits the cosmological constant. The cosmological constant is treated as a fixed parameter in most computations of BH thermodynamics. At at first it has been considered as a dynamical variable in \cite{Henneaux 1984} but later it has been proposed that it is better to suppose as a thermodynamic variable rather than a dynamical variable\cite{Kastor 2009, Doln 2011a, Doln 2011b, Doln 2011c, Jhonson 2011} and is related to the dynamic pressure as :
\begin{equation}\label{ah1.equn1}
P=-\frac{\Lambda}{8\pi}=\frac{3}{8\pi l^2}, 
\end{equation}
where $ \Lambda $ is the cosmological constant, $ l $ is the radius of BHs in AdS (Anti-de Sitter) space and the conjugate quantity of $ P $ is explained as the thermodynamic volume of the BHs \cite{Fan 2016}.

The mass $ M $ of the BHs should be explained as the total gravitational enthalpy $ H $ rather than internal energy of the system in the extended phase apace. In some literature \cite{Kubiznak 2012}, the authors showed that the AdS BH's thermodynamic properties exhibit a number of similarities with Van der Waals (VDW) liquid-gas system if the horizon radius $ r_+ $ of BHs is regarded as the specific volume $ V $. Some authors \cite{Pradhan 2016b} have studied the thermodynamic properties of VDW BHs and relate the thermodynamic volume $ V_t $ of BH with naive geometric volume $ V_g $ and also investigate the thermodynamic stability of BHs by computing specific heat of that BHs.
 
While studying BHs' thermodynamics, the popular approach is to take Bekenstein Entropy. But correction terms in entropy may lead us to different results. Specially, if the BHs are regular, the modification in the definition of entropy may lead us to new results. Enyhalpy calculations for such cases may be interesting.

The paper is organized as follows. In section $2$ we present the basic equations which we use through out the paper. In section $3$ we study the mass of the BHs, surface gravity and temperature of the BHs on event horizon and analyze their behavior graphically and interpret them physically for Hayward class BHs in generic case. We do the same for Asymptotically AdS ( Anti-de Sitter) BHs in section $4$. In section $5$ we examine the stability of both the BHs by calculating the specific heat on event horizon and analyze graphically their behavior and physically interpret them. The logarithmic correction to the entropy for both the BHs are investigated in section $6$. Finally, we conclude this paper in section $7$.     
\section{Basic Equations :}
Assuming that the event horizon of a BHs is a Killing horizon, i.e., that null horizon generators are orbits of a Killing field, the surface gravity  $ \kappa $ of a BH can be defined as the magnitude of the gradient of the norm of horizon generating Killing field $ \chi^a = \zeta^a + \Omega \Psi^a $, evaluated at the horizon. That is, at the horizon, 
\begin{equation}\label{ah1.equn2}
\kappa^2: = -( \nabla^a\vert \Psi\vert)(\nabla_a\vert \Psi\vert)
\end{equation}
Although this surface gravity $ \kappa $ is defined locally on the horizon, it stays constant all over the horizon of a stationary BH. This constancy leads us to draw the resemblance with zeroth law of thermodynamics, which states that the temperature is uniform every where in a system which is in thermal equillibrium. 

Stationary BH solutions of Einstein Field Equation (EFE) relate the first law of thermodynamics as:
\begin{equation}\label{ah1.equn3}
dM=\kappa\frac{dA}{8\pi G} + \Omega dJ + \phi dQ .
\end{equation}\\
Entropy $ S $ is related to the term area $ A $ at event horizon, i.e.,
\begin{equation}\label{ah1.equn4}
S=\frac{A}{4}= \pi r^2 
\end{equation}
as the BH area is given by, 
\begin{equation}\label{ah1.equn5}
 A=\int ^{2\pi}_{0} \int ^{\pi}_{0} \sqrt{g_{\theta \theta}(r_+){g_{\phi\phi}(r_+)}} {d\theta d\phi} = 4\pi r_+^2.
\end{equation}
The classical BH thermodynamics \cite{Bekenstein 1973, Bekenstein 1974, Hawking 1975, Hawking 1976} shows mass $ M $ and surface gravity $ \kappa $ of the BHs related to internal energy $ U $ and  temperature $ T $ as:
\begin{equation}\label{ah1.equn6}
 M=U ~~~~~~~~   and ~~~~~~~~  T=\frac{\kappa}{2\pi}. 
\end{equation}
More generally the mass $ M $ of the BHs is nothing but the total gravitational enthalpy $ H $ \cite{Fan 2016} and is  expressed as:
\begin{equation}\label{ah1.equn7}
M=H=U+PV ,
\end{equation}
where $ P $ is the thermodynamic dynamic pressure and $ V $ is the naive geometric volume and they are defined by
\begin{equation}\label{ah1.equn8}
P=-\frac{\Lambda}{8\pi}=\frac{3}{8\pi l^2} 
~~~~~~~~~ and ~~~~~~~~ V=\frac{4}{3}\pi r_+^3 ,
\end{equation} \\
where $ \Lambda $ is the cosmological constant, $ l $ is the radius of BHs in AdS space and $ r $ is radius of event horizon. Again
\begin{equation}\label{ah1.equn9}
dH=dM=TdS+VdP ,
\end{equation}
which gives,
\begin{equation}\label{ah1.equn10}
V=\left(\frac{\partial H}{\partial P}\right)_S =\left(\frac{\partial M}{\partial P}\right)_S
~~~~~~~~~~~ and~~~~~~~~~~~~T=\left(\frac{\partial H}{\partial S}\right)_P =\left(\frac{\partial M}{\partial S}\right)_P. 
\end{equation}
This volume $ V $ is known as thermodynamic volume and this temperature is the temperature of the BH and it is the same as obtained from the expression $ T=\frac{\kappa}{2\pi} $ but generally thermodynamic volume is not equal to the naive geometric volume of the BHs.
\subsection{Logarithmic Correction of Entropy:}
The partition function is defined as \cite{Gibbons 1976}
\begin{equation}\label{ah1.equn11}
Z(\beta)=\int^{\infty}_{0} \rho(E)e^{-\beta E}dE ,
\end{equation} 
where $ \beta=\frac{1}{T} $ and
$ \rho(E) $ is the density of state and that may be written as an inverse Laplace transformation of the partition function as 
\begin{equation}\label{ah1.equn12}
\rho(E)=\frac{1}{2\pi i} \int^{c+i\infty}_{c-i\infty} Z(\beta)e^{\beta E}d\beta, 
\end{equation}\\
where $ c $ is a constant.
Statistical mechanics defines the entropy of a system as: 
\begin{equation}\label{ah1.equn13}
S=\ln Z + \beta E . 
\end{equation}
Near the equilibrium and inverse temperature $ \beta = \beta_0 $ , the entropy could be expanded as:
\begin{equation}\label{ah1.equn14}
S(\beta)=S(\beta_0) + \frac{1}{2}(\beta - \beta_0)^2 \left.\frac {\partial ^2 S}{\partial \beta ^2} \right|_{\beta = \beta_0} + ........~~~~~~~,
\end{equation}

Hence $ \rho $ is now expressed as:
\begin{equation}\label{ah1.equn15}
\rho (E)=\frac{exp\left\{{S(\beta_0)}\right\}}{2\pi i} \int^{c+i\infty}_{c-i\infty} exp\left\{\frac{(\beta-\beta_{0})^{2} \left(\left.\frac{\partial ^2 S}{\partial \beta ^2} \right|_{\beta = \beta_0}\right)}{2}\right\} d\beta .
\end{equation}
Taking $ \beta - \beta_0 = ix $ 
and $ c = \beta_0 $ where $ x $ is a real variable one can obtain by contour integral that:
\begin{equation}\label{ah1.equn16}
\rho(E)=\frac{exp\left\{{S(\beta_0)}\right\}}{\sqrt{2\pi \left(\left.\frac{\partial ^2 S}{\partial \beta ^2} \right|_{\beta = \beta_0}\right)} } .
\end{equation}
The logarithm of $ \rho (E) $ is nothing but the corrected entropy of the system stated as:
\begin{equation}\label{ah1.equn17}
S= \ln \rho (E)= S(\beta_0)-\frac{1}{2}\ln \left(\left.\frac{\partial ^2 S}{\partial \beta ^2} \right|_{\beta = \beta_0}\right)+......
\end{equation}
\subsection{Thermal Heat Capacity :}
The thermal heat capacity $ C $ is related to the mean value of energy \cite{Sommerfield 1956} in statistical view as:
\begin{equation}\label{ah1.equn18}
C \equiv \left.\frac{\partial\langle H\rangle}{\partial T}\right| _{T=T_0}
= \frac{1}{T^2}\left[ \frac{1}{Z} \left.\frac {\partial ^2 Z}{\partial \beta ^2} \right|_{\beta = \beta_0} - \frac{1}{Z^2 } \left( \frac {\partial Z}{\partial \beta }\right)^2 _{\beta = \beta_0} \right] 
= \frac{\left.\frac {\partial ^2 S}{\partial \beta ^2} \right|_{\beta = \beta_0}}{T^2} ,
\end{equation} 
where the mean value of energy \cite{Sommerfield 1956} is given by:
\begin{equation}\label{ah1.equn19}
\langle E \rangle = -\left.\frac{\partial \left(\ln Z\right)}{\partial \beta}\right| _{\beta = \beta_0} = -\frac{1}{Z} \left.\frac{\partial Z}{\partial \beta}\right| _{\beta = \beta_0}.
\end{equation}
Therefore, the leading order corrections to the BH entropy can be obtained by the relation as:
\begin{equation}\label{ah1.equn20}
S=\ln \rho= S(\beta_0) - \frac{1}{2}\ln \left[CT^2\right] 
\end{equation}
For positive thermal heat capacity or specific heat the $\it{eq^n}$ (\ref{ah1.equn20}) can be written as:
\begin{equation}\label{ah1.equn21}
S=\ln \rho= S(\beta_0) - \frac{1}{2}\ln \left\vert CT^2\right\vert .
\end{equation}
\subsection{Another way to derive the logarithmic correction :}
Exact entropy function $ S(\beta) $, which is followed by conformal field theory (CFT) \cite{Das 2001, Carlip 2000} is depicted as:
\begin{equation}\label{ah1.equn22}
S(\beta)=a\beta+\frac{b}{\beta}.
\end{equation} 
The more general form of the entropy function is
\begin{equation}\label{ah1.equn23}
S(\beta) = a\beta ^m + \frac{b}{\beta^n} ,
\end{equation}
where $ a, b, n, m > 0 $.
The special case, when $ m=n=1 $ and is due to CFT. After some  algebraic calculation one can obtain $ S_0 = \frac{S^"_0}{T^2} $ ( for more detail see \cite{Pradhan 2016a, Das 2001}, then the leading order corrections to the entropy should be as:
\begin{equation}\label{ah1.equn24}
S=\ln\rho = S_0 - \frac{1}{2}\ln\left\vert S_0T^2 \right\vert.
\end{equation}
The beauty of his formula is that one can calculate the entropy of a BH by knowing only the values of $ T $ and $ S_0, $ of that BH. 
\section{Hayward class BH: Generic case :}
The metric of Hayward class BH for Generic case \cite{Fan 2016} can be expressed as:
\begin{equation}\label{ah1.equn25}
ds^2 = -f(r)dt^2 + \frac{dr^2}{f(r)} + r^2 (d\theta^2 + \sin^2 \theta d\phi^2),
\end{equation}
where
\begin{equation}\label{ah1.equn26}
f(r)= 1-\frac{2M}{r}- \frac{2 \alpha^{-1} q^3 r^{\mu -1}}{(r^\nu + q^\nu)^{\frac{\mu}{\nu}}} ,
\end{equation} 
where $ \mu $ and $ \nu $ are both dimensionless constant having value grater then zero, $ \alpha >0 $ has dimension of length squared and $ q \left(= \sqrt{Q_m \sqrt{2\alpha}}\right) $ is a free integrating constant connected with the magnetic charge $ Q_m $. 

To solve the Einstein field equation (EFE) of the form: $ G_{\mu\nu}=8\pi T_{\mu\nu} $ one can get the metric solutions (25).
On the event horizon $ r=r_+ $, mass of this BH can be expressed as:
\begin{equation}\label{ah1.equn27}
M= \frac{r_+}{2}- \frac{\alpha^{-1}q^3 r^{\mu}_+}{(r^\nu_+ + q^\nu)^{\frac{\mu}{\nu}}}.
\end{equation}


We have plotted the mass of BHs $ M $ with $ r_+ $ and $ q $  in Fig-$ 1 $ for $\it{eq^n}$ (\ref{ah1.equn27}). When charge is very low, i.e., when it is similar to Schwarzschild singularity, we see the mass is slowly increasing function of radius of event horizon but as we increase the charge, we see less mass is required to construct a BH for higher radius of event horizon. Normally charge repels the outer region such that more mass is required to bind higher quantity of charge. But in this case we notice the increment in charge forces the BH to be formed at a low mass. This is very particularly is a property of this class of BH.   .\\

The area of the BHs on event horizon $ A_+ $ is expressed as $\it{eq^n}$ (\ref{ah1.equn5}) 
and is related to the total ADM (Arnowitt-Deser-Misner) mass $ M $ of Schwarzschild BH space-time as \cite{Penrose 1973}
\begin{equation}\label{ah1.equn28}
A_+ \leq 16\pi M^2.
\end{equation} \\
This inequality is sometime called `Cosmic Censorship Inequality'.\\  
Using this inequality the mass of this BH can be expressed in terms of $ A_+ $ as:
\begin{equation}\label{ah1.equn29}
M \geq \sqrt{\frac{A_+}{16\pi}} - \frac{\alpha^{-1} q^3 \left( \frac{A_+}{4\pi}\right) ^{\frac{\mu}{2}}}{\left\lbrace \left( \frac{A_+}{4\pi}\right) ^{\frac{\nu}{2}}+ q^{\nu}\right\rbrace ^{\frac{\mu}{\nu}}}.  
\end{equation} \\
The surface gravity of the BH on event horizon is computed as \cite{Bardeen 1973, Bekenstein 1973}:
\begin{equation}\label{ah1.equn30}
\kappa_+ = \frac{f^{'}(r)}{2} = \frac{1}{2r_+} -\frac{\mu \alpha^{-1}q^3r^{\mu- 2}_+}{{(r^\nu_+ + q^\nu)^{\frac{\mu}{\nu}}}} + \frac{\mu \alpha^{-1}q^3r^{\mu + \nu - 2}_+}{{(r^\nu_+ + q^\nu)^{{\frac{\mu}{\nu}}+1}}}.
\end{equation} \\
This quantity is constant over the horizon and that satisfies the zeroth law of thermodynamics.\\
Correspondingly, the temperature of the BH on the horizon could be expressed as:
\begin{equation}\label{ah1.equn31}
T_+ = \frac{1}{4\pi r_+} \left[ 1 -\frac{2\mu \alpha^{-1}q^3r^{\mu- 1}_+}{{(r^\nu_+ + q^\nu)^{\frac{\mu}{\nu}}}} + \frac{2\mu \alpha^{-1}q^3r^{\mu + \nu - 1}_+}{{(r^\nu_+ + q^\nu)^{{\frac{\mu}{\nu}}+1}}}\right]. 
\end{equation}\\
In Fig-$ 2 $, we have plotted the variation of BHs temperature $ T_+$ on event horizon with respect to $ r_+ $ and $ q $ for $\it{eq^n}$ (\ref{ah1.equn31}). Alike the case of Fig-$ 1 $, when charge is very low we observe the temparature is steeply decreasing at low radius of event horizon and is slowly decreasing when we incresae the radius of event horizon. Now as we increase the charge, we see the temparature firstly decreases then after reaching the local minima it starts to increase. Generally occurance of such a local minima signifies a probable transituion of a phase to another. Particularly this may be a sign characteristics for regular BHs  .\\
One can compute the mass of the BH in terms of entropy $ S_+ $ as:
\begin{equation}\label{ah1.equn32}
M = \sqrt{\frac{S_+}{4\pi}} - \frac{\alpha^{-1} q^3 \left( \frac{S_+}{\pi}\right) ^{\frac{\mu}{2}}}{\left\lbrace \left( \frac{S_+}{\pi}\right) ^{\frac{\nu}{2}}+ q^{\nu}\right\rbrace ^{\frac{\mu}{\nu}}}.  
\end{equation} \\


Fig-$ 3 $ represents the variation of mass of the BHs, $ M $ for event horizon with respect to  $ S_+ $ and $ q $ for $\it{eq^n}$ (\ref{ah1.equn32}). In this diagram, when we vary $ M $ with increase of $ S_+ $ keeping $ q $ is fixed, we see that $ M $ increases very slowly with increase of $ S_+ $, but if $ S_+ $ and $ q $ is to be increased simultaneously a curved shape figure is obtained. But its curvature is less than the curvature of Fig-1.\\  
\section{Asymptotically AdS (Anti-de Sitter) BHs:}
Now we take the metric which is the solutions of EFE of the form:  $ G_{\mu\nu}+ \Lambda g_{\mu\nu}= 8\pi T_{\mu\nu} $, where $ \Lambda $ is negative cosmological constant $\it{eq^n}$ (\ref{ah1.equn6}). AdS BH is one of them and the metric in this case is given as \cite{Fan 2016}:
\begin{equation}\label{ah1.equn33}
ds^2 = -f(r)dt^2 + \frac{dr^2}{f(r)} + r^2 (d\theta^2 + \sin^2 \theta d\phi^2),
\end{equation} \\
where
\begin{equation}\label{ah1.equn34}
f(r)= 1-\frac{2M}{r} + \frac{r^2}{l^2}-\frac{2 \alpha^{-1} q^3 r^{\mu -1}}{(r^\nu + q^\nu)^{\frac{\mu}{\nu}}}.
\end{equation} \\
The mass of this BH on the event horizon $ r=r_+ $ is given as:
\begin{equation}\label{ah1.equn35}
M= \frac{r_+}{2}+\frac{r^3_+}{2l^2}- \frac{\alpha^{-1}q^3 r^{\mu}}{(r^\nu_+ + q^\nu)^{\frac{\mu}{\nu}}}.
\end{equation}\\


We have plotted the variation of $ M $ with respect to $ r_+ $ and $ q $  in Fig.s $4a$ and $4b$ for $\it{eq^n}$ (\ref{ah1.equn35}). Fig-$4a $ which plotted for large radius of BHs in AdS space shows the graph whose nature is almost same as Fig-$1$. But if we take the radius of BHs in AdS space very very small, we find that $ M $ increases gradually but not linearly and a cascade feature is obtained under the same, shown in Fig-$4b$. \\\\  
Using the inequality (28) the mass of this BH can be expressed in terms of $ A_+ $ as:
\begin{equation}\label{ah1.equn36}
M \geq \sqrt{\frac{A_+}{16\pi}}+\frac{1}{2 l^2}\left(\frac{A_+}{4\pi}\right)^{\frac{3}{2}} - \frac{\alpha^{-1} q^3 \left( \frac{A_+}{4\pi}\right) ^{\frac{\mu}{2}}}{\left\lbrace \left( \frac{A_+}{4\pi}\right) ^{\frac{\nu}{2}}+ q^{\nu}\right\rbrace ^{\frac{\mu}{\nu}}}.  
\end{equation} \\
This $\it{eq^n}$ (\ref{ah1.equn36}) suggests that the total mass of BH is at least $ \sqrt{\frac{A_+}{16\pi}} $ in a given region of space-time and this inequality signifies that the lower bound on the energy for any time-symmetric data set that agrees the EFE with negative cosmological constant and also coupled to a matter system which possess no necked singularities and satisfies the energy dominant condition.\\  
The surface gravity of the BHs on the event horizon is expressed as:
\begin{equation}\label{ah1.equn37}
\kappa_+ = \frac{f^{'}(r)}{2} = \frac{1}{2r_+}+\frac{r_+}{l^2} -\frac{\mu \alpha^{-1}q^3r^{\mu- 2}_+}{{(r^\nu_+ + q^\nu)^{\frac{\mu}{\nu}}}} + \frac{\mu \alpha^{-1}q^3r^{\mu + \nu - 2}_+}{{(r^\nu_+ + q^\nu)^{{\frac{\mu}{\nu}}+1}}}.
\end{equation} \\
This quantity is constant over the horizon and that satisfies the zeroth law of thermodynamics.\\
Correspondingly, the temperature of the BH on the horizon could be expressed as:
\begin{equation}\label{ah1.equn38}
T_+ = \frac{1}{4\pi r_+} \left[ 1+\frac{2r_+}{l^2} -\frac{2\mu \alpha^{-1}q^3r^{\mu- 1}_+}{{(r^\nu_+ + q^\nu)^{\frac{\mu}{\nu}}}} + \frac{2\mu \alpha^{-1}q^3r^{\mu + \nu - 1}_+}{{(r^\nu_+ + q^\nu)^{{\frac{\mu}{\nu}}+1}}}\right]. 
\end{equation}\\
\vspace{.6cm}


We have plotted the BHs temperature $ T_+ $ on event horizon with $ r_+ $ and $ q $ for $\it{eq^n}$ (\ref{ah1.equn38}). When the variation of $ T_+ $ is to be performed by taking the radius of BHs in AdS space very large (Fig-$ 5a $), we have the similar nature of the curve as Fig-$ 2 $. But if the radius of BHs in AdS space is taken very small, a cascade feature is obtained in the diagram (Fig-$ 5b $) with out changing the basic nature of the curve and it is obvious due to the transition of phases.\\ \\
The mass of the BH in terms of entropy $ S_+ $ and $ P $ is
\begin{equation}\label{ah1.equn39}
M = \sqrt{\frac{S_+}{4\pi}} - \frac{\alpha^{-1} q^3 \left( \frac{S_+}{\pi}\right) ^{\frac{\mu}{2}}}{\left\lbrace \left( \frac{S_+}{\pi}\right) ^{\frac{\nu}{2}}+ q^{\nu}\right\rbrace ^{\frac{\mu}{\nu}}}+\frac{4}{3}P\sqrt{\frac{S_+}{\pi}}.  
\end{equation} \\

We have plotted the variation of BH mass $ M $ with respect to $ S_+ $ and $ q $ for $\it{eq^n}$ (\ref{ah1.equn39}) in  Fig-$ 6 $. In this diagram we show that with increase of $ S_+ $ and $ q $, primarily the mass of BH increases relatively high rate and then the rate of increase becomes low.   \\ \\
Now we examine the naive geometric volume $ V_g $ and thermodynamic volume $ V_t $ of this BH and find out the relation between them.\\

The thermodynamical parameters of this BH on the horizon is obtained from equations (10) and (39) as:
\begin{equation}\label{ah1.equn40}
T_+=\frac{1}{4\sqrt{S_+\pi}}- \frac{\mu\alpha^{-1} q^3 \left( \frac{S_+}{\pi}\right) ^{\frac{\mu}{2}-1}}{2\pi\left\lbrace\left( \frac{S_+}{\pi}\right) ^{\frac{\nu}{2}}+ q^{\nu}\right\rbrace ^{\frac{\mu}{\nu}}}+ \frac{\mu\alpha^{-1} q^3 \left( \frac{S_+}{\pi}\right) ^{\frac{\mu}{2}+\frac{\nu}{2}-1}}{2\pi\left\lbrace\left( \frac{S_+}{\pi}\right) ^{\frac{\nu}{2}}+ q^{\nu}\right\rbrace ^{\frac{\mu}{\nu}+{1}}}+\frac{2}{3}P\sqrt{\frac{1}{S_+\pi}}  
\end{equation} \\

and 
\begin{equation}\label{ah1.equn41}
V_+=\frac{4}{3}\sqrt{\frac{S_+}{\pi}}.
\end{equation}\\
This is actually the thermodynamic volume $ V_t $ of the BH and the navie geometric volume $ V_g $ in terms entropy $ S_+ $ is 
\begin{equation}\label{ah1.equn42}
V_g = \frac{4}{3}S_+\sqrt{\frac{S_+}{\pi}}.
\end{equation}\\ 
We infer from the relations $(41)$ and $(42)$ that for this BH, the entropy on the horizon, $ r=r_+ $  is also defined as the ratio of the naive geometric volume $ V_g $ to the thermodynamic volume $ V_t $.\\\\

In Fig-$ 7 $, we have plotted the variation of BHs temperature $ T_+ $ on event horizon with respect to $ S_+$ and $ q $ for $\it{eq^n}$ (\ref{ah1.equn40}). It is seen from the curve that with increase of $ S_+ $, initially $ T_+ $ decreases rapidly and later the rate of decreases slowly but change in $ q $ does not effect $ T_+ $.\\ 

\section{Thermal Heat Capacity or Specific Heat :}

Now we analyze the local thermodynamic stability of the afore said BHs.
We have to calculate first the thermal heat capacity plainly he specific heat which is given as:
\begin{equation}\label{ah1.equn43}
C_+=\left( \frac{\partial M}{\partial T_+}\right)= \frac{\left( \frac{\partial M}{\partial r_+}\right)}{\left( \frac{\partial T_+}{\partial r_+}\right)} 
\end{equation}\\
and then examine it for different regime.\\
(i) negative specific heat indicates that the BH is thermodynamically unstable.\\
(ii) positive specific heat implies that the BH is thermodynamically stable.\\
(iii) there is also a critical case where the specific heat blows up and that signals a second order phase transition for such BHs.\\\\
\subsection{Hayward class BH: Generic case :}

The specific heat of the Hayward class BHs for Generic case is calculated as
\begin{equation}\label{ah1.equn44}
C_+=-2\pi r^2_+\left[\frac{1-\frac{2\mu \alpha^{-1}q^3r^{\mu- 1}_+}{{(r^\nu_+ + q^\nu)^{\frac{\mu}{\nu}}}} + \frac{2\mu \alpha^{-1}q^3r^{\mu + \nu - 1}_+}{{(r^\nu_+ + q^\nu)^{{\frac{\mu}{\nu}}+1}}}}{1+\frac{2\mu (\mu - 2) \alpha^{-1}q^3r^{\mu- 2}_+}{{(r^\nu_+ + q^\nu)^{\frac{\mu}{\nu}}}} +\frac{2\mu (\mu + \nu) \alpha^{-1}q^3 r^{\mu +2\nu- 3}_+}{{(r^\nu_+ + q^\nu)^{\frac{\mu}{\nu}+2}}}- \frac{2\mu (2\mu + \nu -2)\alpha^{-1}q^3 r^{\mu + \nu - 3}_+}{{(r^\nu_+ + q^\nu)^{{\frac{\mu}{\nu}}+1}}}} \right]. 
\end{equation}\\

\textbf{Case(i)} i.e., $ C_+ <0 $ satisfies when
\begin{equation}\label{ah1.equn45}
1>\frac{2\mu \alpha^{-1}q^3r^{\mu- 1}_+}{{(r^\nu_+ + q^\nu)^{\frac{\mu}{\nu}}}} + \frac{2\mu \alpha^{-1}q^3r^{\mu + \nu - 1}_+}{{(r^\nu_+ + q^\nu)^{{\frac{\mu}{\nu}}+1}}}
\end{equation}\\and
\begin{equation}\label{ah1.equn46}
1+\frac{2\mu (\mu - 2) \alpha^{-1}q^3r^{\mu- 2}_+}{{(r^\nu_+ + q^\nu)^{\frac{\mu}{\nu}}}} +\frac{2\mu (\mu + \nu) \alpha^{-1}q^3 r^{\mu +2\nu- 3}_+}{{(r^\nu_+ + q^\nu)^{\frac{\mu}{\nu}+2}}}> \frac{2\mu (2\mu + \nu -2)\alpha^{-1}q^3 r^{\mu + \nu - 3}_+}{{(r^\nu_+ + q^\nu)^{{\frac{\mu}{\nu}}+1}}}.
\end{equation}\\ 

\textbf{Case(ii)} i.e.,$ C_+ > 0 $ holds good when
\begin{equation}\label{ah1.equn47}
1>\frac{2\mu \alpha^{-1}q^3r^{\mu- 1}_+}{{(r^\nu_+ + q^\nu)^{\frac{\mu}{\nu}}}} + \frac{2\mu \alpha^{-1}q^3r^{\mu + \nu - 1}_+}{{(r^\nu_+ + q^\nu)^{{\frac{\mu}{\nu}}+1}}}
\end{equation}\\ and
\begin{equation}\label{ah1.equn48}
1+\frac{2\mu (\mu - 2) \alpha^{-1}q^3r^{\mu- 2}_+}{{(r^\nu_+ + q^\nu)^{\frac{\mu}{\nu}}}} +\frac{2\mu (\mu + \nu) \alpha^{-1}q^3 r^{\mu +2\nu- 3}_+}{{(r^\nu_+ + q^\nu)^{\frac{\mu}{\nu}+2}}} 1< \frac{2\mu (2\mu + \nu -2)\alpha^{-1}q^3 r^{\mu + \nu - 3}_+}{{(r^\nu_+ + q^\nu)^{{\frac{\mu}{\nu}}+1}}}
\end{equation}\\ or
\begin{equation}\label{ah1.equn49}
1 < \frac{2\mu \alpha^{-1}q^3r^{\mu- 1}_+}{{(r^\nu_+ + q^\nu)^{\frac{\mu}{\nu}}}} + \frac{2\mu \alpha^{-1}q^3r^{\mu + \nu - 1}_+}{{(r^\nu_+ + q^\nu)^{{\frac{\mu}{\nu}}+1}}}
\end{equation}\\ and
\begin{equation}\label{ah1.equn50}
1+\frac{2\mu (\mu - 2) \alpha^{-1}q^3r^{\mu- 2}_+}{{(r^\nu_+ + q^\nu)^{\frac{\mu}{\nu}}}} +\frac{2\mu (\mu + \nu) \alpha^{-1}q^3 r^{\mu +2\nu- 3}_+}{{(r^\nu_+ + q^\nu)^{\frac{\mu}{\nu}+2}}}> \frac{2\mu (2\mu + \nu -2)\alpha^{-1}q^3 r^{\mu + \nu - 3}_+}{{(r^\nu_+ + q^\nu)^{{\frac{\mu}{\nu}}+1}}}.
\end{equation}\\ 

\textbf{Case(iii)} i.e., $ C_+\rightarrow \infty $ , the critical condition agrees when
\begin{equation}\label{ah1.equn51}
1+\frac{2\mu (\mu - 2) \alpha^{-1}q^3r^{\mu- 2}_+}{{(r^\nu_+ + q^\nu)^{\frac{\mu}{\nu}}}} +\frac{2\mu (\mu + \nu) \alpha^{-1}q^3 r^{\mu +2\nu- 3}_+}{{(r^\nu_+ + q^\nu)^{\frac{\mu}{\nu}+2}}}= \frac{2\mu (2\mu + \nu -2)\alpha^{-1}q^3 r^{\mu + \nu - 3}_+}{{(r^\nu_+ + q^\nu)^{{\frac{\mu}{\nu}}+1}}}.
\end{equation}\\

\begin{figure}[h!]
\begin{center}
\includegraphics[scale=.8]{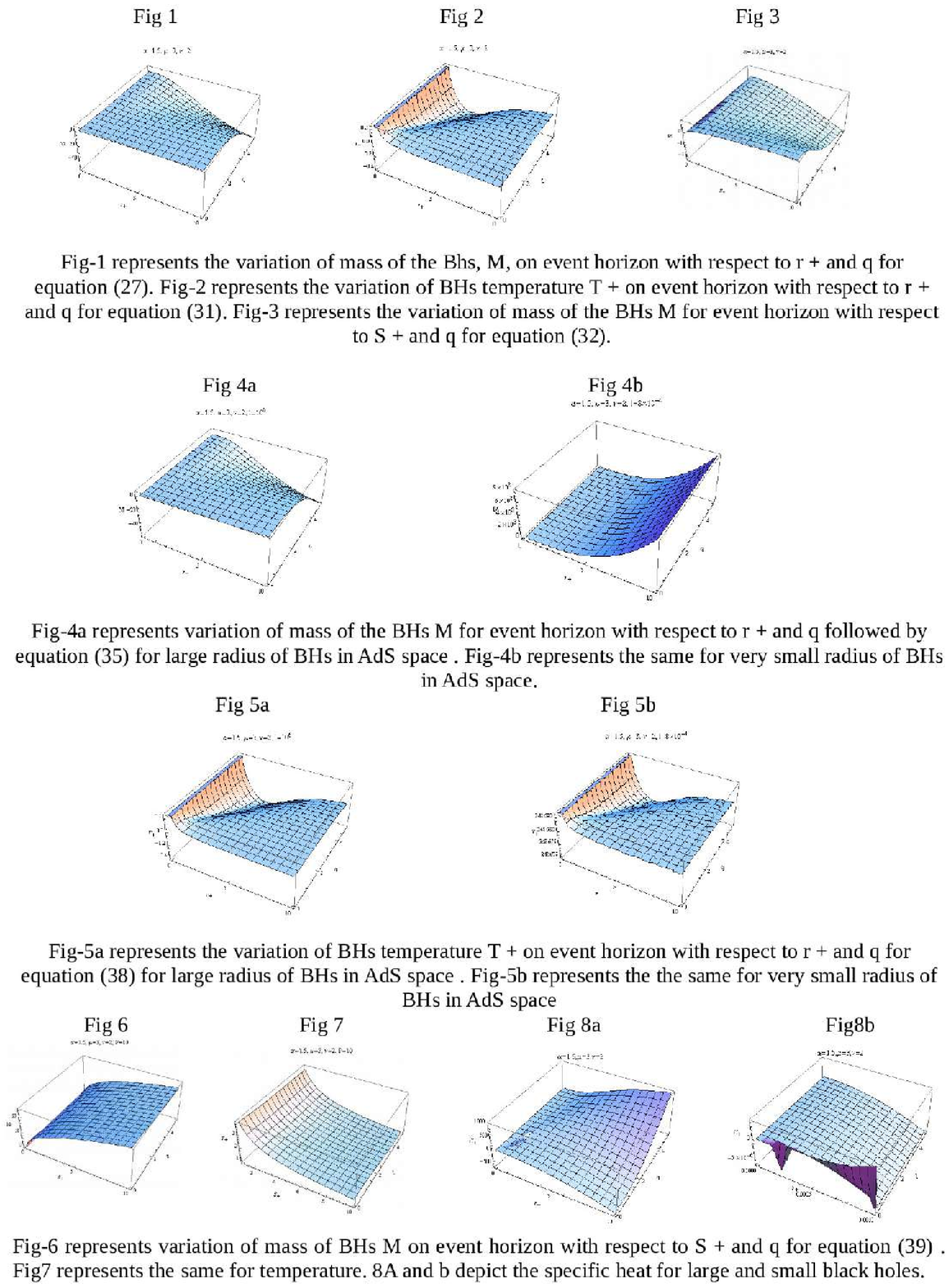}~~~~
\end{center}
\end{figure}.\\
If we vary the specific heat of BHs $ C_+ $ on event horizon with $ r_+ $ and $ q $ for $\it{eq^n}$ (\ref{ah1.equn44}) for large radius of event horizon we obtain the curve as Fig-$ 8a $. In this diagram we follow a little peak which is obvious for the transition of phases for the low values of radius of event horizon. Due to increase of $ r_+ $, $ C_+ $ decreases keeping $ q $ is constant and $ C_+ $ increases as $ q $ increases. Hence due to increase of both $ r_+$ and $ q $ simultaneously, we get the feature of the graph as Fig-$ 8a $. In Fig-$ 8b $, we have showed the variation of specific heat $ C_+ $ on event horizon with respect to $ r_+ $ and $ q $ for $\it{eq^n}$ (\ref{ah1.equn44}) for very small radius of event horizon. Here we observe when $ r_+ $ very near to zero a sudden decrease of $ C_+ $ occurs and later slight increase of $ r_+ $, $ C_+ $ increases highly rapid to a certain extent. Moreover in this region of $r_+$, $q$ dose not effect $C_+$.  
\subsection{Asymptotically AdS (Anti-de Sitter) BHs :} 

The specific heat of Asymptotically AdS BHs are computed as:
\begin{equation}\label{ah1.equn52}
C_+=-2\pi r^2_+\left[\frac{1+ \frac{2r_+}{l^2}-\frac{2\mu \alpha^{-1}q^3r^{\mu- 1}_+}{{(r^\nu_+ + q^\nu)^{\frac{\mu}{\nu}}}} + \frac{2\mu \alpha^{-1}q^3r^{\mu + \nu - 1}_+}{{(r^\nu_+ + q^\nu)^{{\frac{\mu}{\nu}}+1}}}}{1 - \frac{2r^2_+}{l^2}+\frac{2\mu (\mu - 2) \alpha^{-1}q^3r^{\mu- 2}_+}{{(r^\nu_+ + q^\nu)^{\frac{\mu}{\nu}}}} +\frac{2\mu (\mu + \nu) \alpha^{-1}q^3 r^{\mu +2\nu- 3}_+}{{(r^\nu_+ + q^\nu)^{\frac{\mu}{\nu}+2}}}- \frac{2\mu (2\mu + \nu -2)\alpha^{-1}q^3 r^{\mu + \nu - 3}_+}{{(r^\nu_+ + q^\nu)^{{\frac{\mu}{\nu}}+1}}}} \right] .
\end{equation}\\
\begin{figure}[h!]
\begin{center}

\includegraphics[scale=.8]{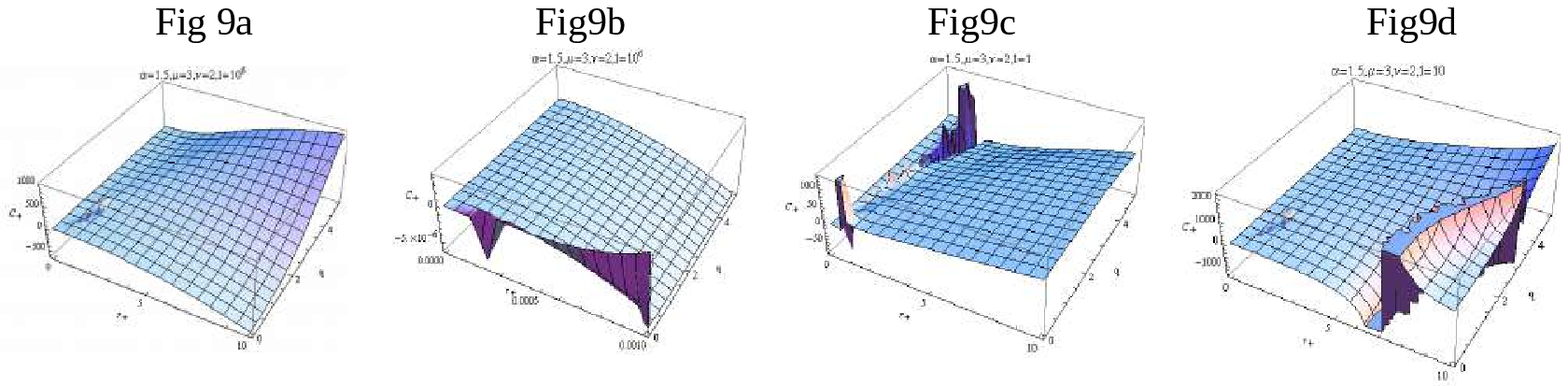}~~~~


\end{center}
\end{figure}\\
As for large values of $ l $, $\it{eq^n}$ (\ref{ah1.equn52}) reduces to $\it{eq^n}$ (\ref{ah1.equn44}) and we get almost same features of the graph as Fig.s-$ 8a $ and $ 8b $ when we vary $ C_+ $ with $ r_+ $ and $ q $ (Fig-$ 9a $ and Fig-$ 9b $). But when we vary $ C_+ $ with same variable only taking the values of $ l $ small, we observe very surprising features of the curves ( Fig-$ 9c $ and Fig-$ 9d $). In Fig-$ 9c $ we have plotted $ C_+ $ with $ r_+ $ and $ q $ taking $ l=1 $. Here we find a sharp peak for low values of $ r_+ $ and when $ q $ increases with same values of $ r_+ $ a discontinuous pattern is occurred and further increases of $ r_+ $, $ C_+ $ decreases relatively greater than that due to increases of $ q $. When we take $ l=10 $, we obtain Fig-$ 9d $. In this figure we find initially $ C_+ $ decreases smoothly for some extent with increase of $ r_+ $ and suddenly occurs a high peak, with increase of $ q $, primarily this peak falls with sudden and then increases gradually with further increase of $ q $. All these peaks indicate that there happen the transitions of phases.      

\textbf{Case(i)} i.e., $ C_+ <0 $ satisfies when
\begin{equation}\label{ah1.equn53}
1> \frac{2r_+}{l^2}+\frac{2\mu \alpha^{-1}q^3r^{\mu- 1}_+}{{(r^\nu_+ + q^\nu)^{\frac{\mu}{\nu}}}} + \frac{2\mu \alpha^{-1}q^3r^{\mu + \nu - 1}_+}{{(r^\nu_+ + q^\nu)^{{\frac{\mu}{\nu}}+1}}}
\end{equation}\\and
\begin{equation}\label{ah1.equn54}
1+\frac{2\mu (\mu - 2) \alpha^{-1}q^3r^{\mu- 2}_+}{{(r^\nu_+ + q^\nu)^{\frac{\mu}{\nu}}}} +\frac{2\mu (\mu + \nu) \alpha^{-1}q^3 r^{\mu +2\nu- 3}_+}{{(r^\nu_+ + q^\nu)^{\frac{\mu}{\nu}+2}}}> \frac{2\mu (2\mu + \nu -2)\alpha^{-1}q^3 r^{\mu + \nu - 3}_+}{{(r^\nu_+ + q^\nu)^{{\frac{\mu}{\nu}}+1}}}+\frac{2r^2_+}{l^2}.
\end{equation}\\ 
\textbf{Case(ii)} i.e., $ C_+ > 0 $ holds good when
\begin{equation}\label{ah1.equn55}
1> \frac{2r_+}{l^2}+\frac{2\mu \alpha^{-1}q^3r^{\mu- 1}_+}{{(r^\nu_+ + q^\nu)^{\frac{\mu}{\nu}}}} + \frac{2\mu \alpha^{-1}q^3r^{\mu + \nu - 1}_+}{{(r^\nu_+ + q^\nu)^{{\frac{\mu}{\nu}}+1}}}
\end{equation}\\and
\begin{equation}\label{ah1.equn56}
1+\frac{2\mu (\mu - 2) \alpha^{-1}q^3r^{\mu- 2}_+}{{(r^\nu_+ + q^\nu)^{\frac{\mu}{\nu}}}} +\frac{2\mu (\mu + \nu) \alpha^{-1}q^3 r^{\mu +2\nu- 3}_+}{{(r^\nu_+ + q^\nu)^{\frac{\mu}{\nu}+2}}} < \frac{2\mu (2\mu + \nu -2)\alpha^{-1}q^3 r^{\mu + \nu - 3}_+}{{(r^\nu_+ + q^\nu)^{{\frac{\mu}{\nu}}+1}}}+\frac{2r^2_+}{l^2}
\end{equation}\\ or
\begin{equation}\label{ah1.equn57}
1 < \frac{2r_+}{l^2}+\frac{2\mu \alpha^{-1}q^3r^{\mu- 1}_+}{{(r^\nu_+ + q^\nu)^{\frac{\mu}{\nu}}}} + \frac{2\mu \alpha^{-1}q^3r^{\mu + \nu - 1}_+}{{(r^\nu_+ + q^\nu)^{{\frac{\mu}{\nu}}+1}}}
\end{equation}\\and
\begin{equation}\label{ah1.equn58}
1+\frac{2\mu (\mu - 2) \alpha^{-1}q^3r^{\mu- 2}_+}{{(r^\nu_+ + q^\nu)^{\frac{\mu}{\nu}}}} +\frac{2\mu (\mu + \nu) \alpha^{-1}q^3 r^{\mu +2\nu- 3}_+}{{(r^\nu_+ + q^\nu)^{\frac{\mu}{\nu}+2}}}> \frac{2\mu (2\mu + \nu -2)\alpha^{-1}q^3 r^{\mu + \nu - 3}_+}{{(r^\nu_+ + q^\nu)^{{\frac{\mu}{\nu}}+1}}}+\frac{2r^2_+}{l^2}.
\end{equation}\\ 
\textbf{Case(iii)} i.e., $ C_+\rightarrow \infty $ , the critical condition satisfies   when
\begin{equation}\label{ah1.equn59}
1+\frac{2\mu (\mu - 2) \alpha^{-1}q^3r^{\mu- 2}_+}{{(r^\nu_+ + q^\nu)^{\frac{\mu}{\nu}}}} +\frac{2\mu (\mu + \nu) \alpha^{-1}q^3 r^{\mu +2\nu- 3}_+}{{(r^\nu_+ + q^\nu)^{\frac{\mu}{\nu}+2}}} = \frac{2\mu (2\mu + \nu -2)\alpha^{-1}q^3 r^{\mu + \nu - 3}_+}{{(r^\nu_+ + q^\nu)^{{\frac{\mu}{\nu}}+1}}}+\frac{2r^2_+}{l^2}.
\end{equation}\\

\section{Logarithmic Correction to the Entropy :}
\subsection{Hayward class BH: Generic case :}
Now we obtain the logarithmic correction to this BH entropy on the horizon by applying 
the $\it{eq^n}$ (\ref{ah1.equn21}) as :
\begin{equation}\label{ah1.equn60}
S^c_+ = \pi r^2_+ - \frac{1}{2}ln \left\vert \frac{\left[ 1 -\frac{2\mu \alpha^{-1}q^3r^{\mu- 1}_+}{{(r^\nu + q^\nu)^{\frac{\mu}{\nu}}}} + \frac{2\mu \alpha^{-1}q^3r^{\mu + \nu - 1}_+}{{(r^\nu + q^\nu)^{{\frac{\mu}{\nu}}+1}}}\right]^3 }{8\pi\left[{1+\frac{2\mu (\mu - 2) \alpha^{-1}q^3r^{\mu- 2}_+}{{(r^\nu_+ + q^\nu)^{\frac{\mu}{\nu}}}} +\frac{2\mu (\mu + \nu) \alpha^{-1}q^3 r^{\mu +2\nu- 3}_+}{{(r^\nu_+ + q^\nu)^{\frac{\mu}{\nu}+2}}}- \frac{2\mu (2\mu + \nu -2)\alpha^{-1}q^3 r^{\mu + \nu - 3}_+}{{(r^\nu_+ + q^\nu)^{{\frac{\mu}{\nu}}+1}}}}\right] }\right\vert+....
\end{equation}\\
This is the corrected microcanonical entropy due to quantum thermal fluctuations around the equilibrium and it is only valid in the  regime when 
\begin{equation}\label{ah1.equn61}
 1 > \frac{2\mu \alpha^{-1}q^3r^{\mu- 1}_+}{{(r^\nu_+ + q^\nu)^{\frac{\mu}{\nu}}}} + \frac{2\mu \alpha^{-1}q^3r^{\mu + \nu - 1}_+}{{(r^\nu_+ + q^\nu)^{{\frac{\mu}{\nu}}+1}}}
\end{equation} and 
\begin{equation}\label{ah1.equn62}
1+\frac{2\mu (\mu - 2) \alpha^{-1}q^3r^{\mu- 2}_+}{{(r^\nu_+ + q^\nu)^{\frac{\mu}{\nu}}}} +\frac{2\mu (\mu + \nu) \alpha^{-1}q^3 r^{\mu +2\nu- 3}_+}{{(r^\nu_+ + q^\nu)^{\frac{\mu}{\nu}+2}}} > \frac{2\mu (2\mu + \nu -2)\alpha^{-1}q^3 r^{\mu + \nu - 3}_+}{{(r^\nu_+ + q^\nu)^{{\frac{\mu}{\nu}}+1}}}.
\end{equation}\\
Further we calculate the logarithmic correction to the BH entropy by applying Eq.(24)as:
\begin{equation}\label{ah1.equn63}
S^{cft}_+ = \pi r^2_+ - \frac{1}{2}ln \left\vert \frac{\left[ 1 -\frac{2\mu \alpha^{-1}q^3r^{\mu- 1}_+}{{(r^\nu_+ + q^\nu)^{\frac{\mu}{\nu}}}} + \frac{2\mu \alpha^{-1}q^3r^{\mu + \nu - 1}_+}{{(r^\nu_+ + q^\nu)^{{\frac{\mu}{\nu}}+1}}}\right]^2 }{16\pi}\right\vert+....
\end{equation}\\
The numerical value which will get from $\it{eq^n}$ (\ref{ah1.equn60}) and $\it{eq^n}$ (\ref{ah1.equn63}) slightly different due existences of specific heat in $\it{eq^n}$ (\ref{ah1.equn60}) and entropy in $\it{eq^n}$ (\ref{ah1.equn63}) on the horizon. $ S^{cft}_+$ (63) is valid in the domain when
\begin{equation}\label{ah1.equn64}
1>\frac{2\mu \alpha^{-1}q^3r^{\mu- 1}_+}{{(r^\nu_+ + q^\nu)^{\frac{\mu}{\nu}}}} + \frac{2\mu \alpha^{-1}q^3r^{\mu + \nu - 1}_+}{{(r^\nu_+ + q^\nu)^{{\frac{\mu}{\nu}}+1}}}
\end{equation}\\
Now we compare the results given from $\it{eq^n}$ (\ref{ah1.equn60}) and $\it{eq^n}$ (\ref{ah1.equn63}). Before logarithmic correction to entropy of BHs the ratio of $ S^c_+ $ and $ S^{cft}_+ $ is unity i.e.,
\begin{equation}\label{ah1.equn65}
\frac{S^c_+}{S^{cft}_+}=1
\end{equation} which is as expected.
But we have seen an interesting result after taking the logarithmic correction to said entropy that 
\begin{equation}\label{ah1.equn66}
\frac{S^c_+}{S^{cft}_+}= \frac{1- \frac{1}{2\pi r^2_+}ln \left\vert \frac{\left[ 1 -\frac{2\mu \alpha^{-1}q^3r^{\mu- 1}_+}{{(r^\nu_+ + q^\nu)^{\frac{\mu}{\nu}}}} + \frac{2\mu \alpha^{-1}q^3r^{\mu + \nu - 1}_+}{{(r^\nu_+ + q^\nu)^{{\frac{\mu}{\nu}}+1}}}\right]^3 }{8\pi\left[{1+\frac{2\mu (\mu - 2) \alpha^{-1}q^3r^{\mu- 2}_+}{{(r^\nu_+ + q^\nu)^{\frac{\mu}{\nu}}}} +\frac{2\mu (\mu + \nu) \alpha^{-1}q^3 r^{\mu +2\nu- 3}_+}{{(r^\nu_+ + q^\nu)^{\frac{\mu}{\nu}+2}}}- \frac{2\mu (2\mu + \nu -2)\alpha^{-1}q^3 r^{\mu + \nu - 3}_+}{{(r^\nu_+ + q^\nu)^{{\frac{\mu}{\nu}}+1}}}}\right] }\right\vert+....}{1 - \frac{1}{2\pi r^2_+}ln \left\vert \frac{\left[ 1 -\frac{2\mu \alpha^{-1}q^3r^{\mu- 1}_+}{{(r^\nu_+ + q^\nu)^{\frac{\mu}{\nu}}}} + \frac{2\mu \alpha^{-1}q^3r^{\mu + \nu - 1}_+}{{(r^\nu_+ + q^\nu)^{{\frac{\mu}{\nu}}+1}}}\right]^2 }{16\pi}\right\vert+....}
\end{equation}\\
It is very difficult to say whether this ratio smaller or greater than that of unity.\\
\subsection{Asymptotically AdS (Anti-de Sitter) BHs :}

Now we obtain the logarithmic correction to this BH entropy on the horizon by applying the $\it{eq^n}$ (\ref{ah1.equn22}) as :
\begin{equation}\label{ah1.equn67}
S^c_+ = \pi r^2_+ - \frac{1}{2}ln \left\vert \frac{\left[ 1+\frac{2r_+}{l^2} -\frac{2\mu \alpha^{-1}q^3r^{\mu- 1}_+}{{(r^\nu + q^\nu)^{\frac{\mu}{\nu}}}} + \frac{2\mu \alpha^{-1}q^3r^{\mu + \nu - 1}_+}{{(r^\nu + q^\nu)^{{\frac{\mu}{\nu}}+1}}}\right]^3 }{8\pi\left[{1-\frac{2r^2_+}{l^2} +\frac{2\mu (\mu - 2) \alpha^{-1}q^3r^{\mu- 2}_+}{{(r^\nu_+ + q^\nu)^{\frac{\mu}{\nu}}}} +\frac{2\mu (\mu + \nu) \alpha^{-1}q^3 r^{\mu +2\nu- 3}_+}{{(r^\nu_+ + q^\nu)^{\frac{\mu}{\nu}+2}}}- \frac{2\mu (2\mu + \nu -2)\alpha^{-1}q^3 r^{\mu + \nu - 3}_+}{{(r^\nu_+ + q^\nu)^{{\frac{\mu}{\nu}}+1}}}}\right] }\right\vert+....
\end{equation}\\
This is the corrected microcanonical entropy due to quantum thermal fluctuations around the equilibrium and it is only valid in the  regime when 
\begin{equation}\label{ah1.equn68}
 1+\frac{2r_+}{l^2} > \frac{2\mu \alpha^{-1}q^3r^{\mu- 1}_+}{{(r^\nu_+ + q^\nu)^{\frac{\mu}{\nu}}}} + \frac{2\mu \alpha^{-1}q^3r^{\mu + \nu - 1}_+}{{(r^\nu_+ + q^\nu)^{{\frac{\mu}{\nu}}+1}}}
\end{equation} and 
\begin{equation}\label{ah1.equn69}
1+\frac{2\mu (\mu - 2) \alpha^{-1}q^3r^{\mu- 2}_+}{{(r^\nu_+ + q^\nu)^{\frac{\mu}{\nu}}}} +\frac{2\mu (\mu + \nu) \alpha^{-1}q^3 r^{\mu +2\nu- 3}_+}{{(r^\nu_+ + q^\nu)^{\frac{\mu}{\nu}+2}}} > \frac{2\mu (2\mu + \nu -2)\alpha^{-1}q^3 r^{\mu + \nu - 3}_+}{{(r^\nu_+ + q^\nu)^{{\frac{\mu}{\nu}}+1}}}+\frac{2r^2_+}{l^2}
\end{equation}\\
Further we calculate the logarithmic correction to the BH entropy by applying Eq.(24)as:
\begin{equation}\label{ah1.equn70}
S^{cft}_+ = \pi r^2_+ - \frac{1}{2}ln \left\vert \frac{\left[ 1+\frac{2r_+}{l^2} -\frac{2\mu \alpha^{-1}q^3r^{\mu- 1}_+}{{(r^\nu_+ + q^\nu)^{\frac{\mu}{\nu}}}} + \frac{2\mu \alpha^{-1}q^3r^{\mu + \nu - 1}_+}{{(r^\nu_+ + q^\nu)^{{\frac{\mu}{\nu}}+1}}}\right]^2 }{16\pi}\right\vert+....
\end{equation}\\
The numerical value which will get from $\it{eq^n}$ (\ref{ah1.equn67}) and $\it{eq^n}$ (\ref{ah1.equn70}) slightly different due existences of specific heat in $\it{eq^n}$ (\ref{ah1.equn67}) and entropy in $\it{eq^n}$ (\ref{ah1.equn70}) on the horizon. $ S^{cft}_+$ (70) is valid in the domain when
\begin{equation}\label{ah1.equn71}
1+\frac{2r_+}{l^2}>\frac{2\mu \alpha^{-1}q^3r^{\mu- 1}_+}{{(r^\nu_+ + q^\nu)^{\frac{\mu}{\nu}}}} + \frac{2\mu \alpha^{-1}q^3r^{\mu + \nu - 1}_+}{{(r^\nu_+ + q^\nu)^{{\frac{\mu}{\nu}}+1}}}
\end{equation}\\
Now we compare the results given from $\it{eq^n}$ (\ref{ah1.equn67}) and $\it{eq^n}$ (\ref{ah1.equn70}). Before logarithmic correction to entropy of BH it is found that the ratio 
\begin{equation}\label{ah1.equn72}
\frac{S^c_+}{S^{cft}_+}=1
\end{equation} i.e., two entropies are same and which is as expected.
But we have seen an interesting result if we have taken the logarithmic correction to said entropy that 
\begin{equation}\label{ah1.equn73} 
\frac{S^c_+}{S^{cft}_+}=\frac{1- \frac{1}{2\pi r^2_+}ln \left\vert \frac{\left[ 1+\frac{2r_+}{l^2} -\frac{2\mu \alpha^{-1}q^3r^{\mu- 1}_+}{{(r^\nu_+ + q^\nu)^{\frac{\mu}{\nu}}}} + \frac{2\mu \alpha^{-1}q^3r^{\mu + \nu - 1}_+}{{(r^\nu_+ + q^\nu)^{{\frac{\mu}{\nu}}+1}}}\right]^3 }{8\pi\left[{1-\frac{2r^2_+}{l^2}+\frac{2\mu (\mu - 2) \alpha^{-1}q^3r^{\mu- 2}_+}{{(r^\nu_+ + q^\nu)^{\frac{\mu}{\nu}}}} +\frac{2\mu (\mu + \nu) \alpha^{-1}q^3 r^{\mu +2\nu- 3}_+}{{(r^\nu_+ + q^\nu)^{\frac{\mu}{\nu}+2}}}- \frac{2\mu (2\mu + \nu -2)\alpha^{-1}q^3 r^{\mu + \nu - 3}_+}{{(r^\nu_+ + q^\nu)^{{\frac{\mu}{\nu}}+1}}}}\right] }\right\vert+....}{1 - \frac{1}{2\pi r^2_+}ln \left\vert \frac{\left[ 1+\frac{2r_+}{l^2} -\frac{2\mu \alpha^{-1}q^3r^{\mu- 1}_+}{{(r^\nu_+ + q^\nu)^{\frac{\mu}{\nu}}}} + \frac{2\mu \alpha^{-1}q^3r^{\mu + \nu - 1}_+}{{(r^\nu_+ + q^\nu)^{{\frac{\mu}{\nu}}+1}}}\right]^2 }{16\pi}\right\vert+....}
\end{equation}\\
It is very difficult to say whether this ratio smaller or greater than that of unity.\\\\
\section{Conclusion}
In our study, we have investigated the thermodynamic properties like surface gravity $ \kappa $, temperature $ T $ on event horizon ($ r=r_+ $) of regular BHs viz. `Hayward Class and asymptotically AdS BHs' separately. We have plotted different graphs and try to interpret them physically. In Fig-$ 1 $, we have got an interesting feature of the curve that mass $ M $ of Hayward Class BHs slightly increase with increase of radius of event horizon $ r_+ $ keeping $ q $ is constant, but when $ r_+ $ and $ q $ increase simultaneously, $ M $ decreases and a curved shape is obtained. Again for asymptotically AdS BHs  we have shown almost same nature of the curve as Fig-$ 1 $ for large values of radius of BHs in AdS space $ l $ for the same conditions (Fig-$4a$) but when we study it for very small values of $ l $, we find $ M $ increases gradually under the same condition not only that but also there occurs a cascade feature in the diagram (Fig-$4b$) . When we have plotted $ M $ with BHs entropy on event horizon $ S_+ $ and $ q $, we obtain almost the same nature of the curve as Fig-$ 1 $ for Hayward class BH (Fig-$ 3 $) but for asymptotically AdS BHs with increase of $S_+$, initially $M$ increases with high rate and later the rate of increase becomes low (Fig-$ 6 $). We showed almost similar feature of the curves when we plot the BHs temperature on the horizon $ T_+ $ with $ r_+$ and $ q $ for both the `Hayward Class and asymptotically AdS BHs (for both large and very small radii of BHs in the AdS space $ l $). Here we find a phase transition which occurs at low radius of event horizon (Fig-$ 2 $, Fig-$ 5a $ and Fig-$5b$). If the radius of BHs in the AdS space $ l $ very low a cascade feature is occurred in the diagram (Fig-$4b$ and Fig-$ 5b $) keeping the basic nature of the curve unaltered and it is obvious due to the transition of phases. In the variation of $ T_+ $ with respect to $ S_+ $ and $ q $ we have not observed any cascade feature (Fig-$ 7 $). Here we have showed a gradually decreasing curve due to increase of $ S_+$ and $ q $, though the rate of decreasing is not same all over the region. It is initially high and later slows down. When we have varried the specific heat of BHs $ C_+ $ on event horizon with $ r_+ $ and $ q $ for Hayward class BHs for large radius of event horizon we have obtained the curve as Fig-$ 8a $. For the low values of radius of event horizon we followed a little peak which is obvious for the transition of phases. We have computed the same plot for low radius of event horizon in Fig-$ 8b $ further and we have found how the phase transition occurs clearly. As for large values of $ l $ for  asymptotically AdS BHs the variation of the specific heat of BHs $ C_+ $ on event horizon with $ r_+ $ and $ q $ is same due to approximation and so we get almost same features of the curve as Fig.s-$ 8a $ and $ 8b $ (Fig-$ 9a $ and Fig-$ 9b $). We have observed very interesting features of the curves (Fig-$ 9c $ and Fig-$ 9d $), if we vary $ C_+ $ with same variable only taking the values of $ l $ small. For $ l=1 $ we have observed a sharp peak occurs for low values of $ r_+ $ and when $ q $ increases with same values of $ r_+ $ a discontinuous pattern is occurred and further increases of $ r_+ $, $ C_+ $ decreases relatively greater than that due to increases of $ q $. Again if we take $ l=10 $, we have found initially $ C_+ $ decreases smoothly for some extent with increase of $ r_+ $ and suddenly occurs a high peak, with increase of $ q $ primarily this peak falls with sudden and then increases gradually with further increase of $ q $ (Fig-$ 9d $). All these peaks indicates that there occurs the transition of phases.\\
In \cite{Pradhan 2016b}, a similar work has been done for Van-der-Waals (VDW) BHs and here the author have showed  that the thermodynamic volume is greater than that of the naive geometric volume of that BHs. But for asymptotically AdS BHs we found that the entropy on the event horizon is simply the ratio of naive geometric volume to thermodynamic volume.\\
We have derived the `cosmic-Censorship-Inequality' for both the BHs. Moreover we have calculated the thermal heat capacity of afore said BHs and studied their stability in different regime. Finally we have computed the logarithmic correction to the entropy for both the BHs considering the quantum fluctuations around the thermal equilibrium.

\vspace{.1 in}
{\bf Acknowledgment:}
AH wishes to thank the Department of Mathematics, the University of Burdwan for the research facilities provided during the work. RB thanks IUCAA, PUNE for Visiting Associateship.

\end{document}